# Electronic Structure of the Dilute Magnetic Semiconductor Ga₁₋ₓMnₓP from Hard X-ray Photoelectron Spectroscopy and Hard X-ray Angle-Resolved Photoemission

A. Keqi[1, 2], M. Gehlmann[1, 2, 3], G. Conti[1, 2], S. Nemšák[1, 2, 3], A. Rattanachata[1, 2],

J. Minár[4], L. Plucinski[3], J. E. Rault[5], J. P. Rueff[5], M. Scarpulla[2, 6, 7], M. Hategan[1], G. K.

Pálsson[1, 2, 8], C. Conlon[1, 2], D. Eiteneer[1, 2], A. Y. Saw[1, 2], A. X. Gray[9], K. Kobayashi[10],

S. Ueda[11,12], O. D. Dubon[2, 6], C.M. Schneider[1, 2, 3], C. S. Fadley[1, 2]

*[1]Department of Physics, University of California Davis, Davis, California 95616, U.S.A*

*[2]Materials Sciences Division, Lawrence Berkeley National Laboratory,*

*Berkeley, California 94720, U.S.A*

*[3]Peter-Grünberg-Institut PGI-6, Forschungszentrum Jülich, 52425 Jülich, Germany*

*[4] New Technologies-Research Center, University of West Bohemia,*

*306 14 Pilsen, Czech Republic*

*[5]Synchrotron SOLEIL, Saint-Aubin, 91192, France*

*[6]Department of Material Science and Engineering, University of California Berkeley,*

*Berkeley, California 94720, U.S.A*

*[7]Department of Electrical and Computer Engineering, University of Utah,*

*Salt Lake City, Utah 84112, U.S.A*

*[8]Department of Physics, Uppsala University, Uppsala SE-751 20, Sweden*

*[9]Department of Physics, Temple University, Philadelphia, Pennsylvania 19122, U.S.A*

[10] *Japan Atomic Energy Agency, 1-1-1 Kouto, Sayo-cho, Sayo-gun, Hyogo 679-5148, Japan*

[11] *Synchrotron X-ray Station at SPring-8, National Institute for Materials Science (NIMS), 1-1-1 Kouto, Sayo-cho, Sayo-gun, Hyogo 679-5148, Japan*

[12]*Research Center for Advanced Measurement and Characterization, NIMS, 1-2-1 Sengen, Tsukuba, Ibaraki 305-0047, Japan*

## ABSTRACT

We have investigated the electronic structure of the dilute magnetic semiconductor (DMS) $Ga_{0.98}Mn_{0.02}P$ and compared it to that of an undoped GaP reference sample, using hard X-ray photoelectron spectroscopy (HXPS) and hard X-ray angle-resolved photoemission spectroscopy (HARPES) at energies of about 3 keV. We present experimental data, as well as theoretical calculations, in order to understand the role of the Mn dopant in the emergence of ferromagnetism in this material. Both core-level spectra and angle-resolved or angle-integrated valence spectra are discussed. In particular, the HARPES experimental data are compared to free-electron final-state model calculations and to more accurate one-step photoemission theory. The experimental results show differences between $Ga_{0.98}Mn_{0.02}P$ and GaP in both angle-resolved and angle-integrated valence spectra. The $Ga_{0.98}Mn_{0.02}P$ bands are broadened due to the presence of Mn impurities that disturb the long-range translational order of the host GaP crystal. Mn-induced changes of the electronic structure are observed over the entire valence band range, including the presence of a distinct impurity band close to the valence-band maximum of the DMS. These experimental results are in good agreement with the one-step photoemission calculations, and a prior HARPES study of $Ga_{0.97}Mn_{0.03}As$ and GaAs (Gray et al. Nature Materials 11, 957 (2012)), demonstrating the strong similarity between these two materials. The Mn $2p$ and $3s$ core-level spectra also reveal an essentially identical state in doping both GaAs and GaP.

## I. Introduction

Since their first synthesis and discussion[1, 2] *dilute magnetic semiconductors* (DMSs) have been studied widely[3, 4]. These materials have potential applications in *spin transport electronics* (spintronics), e.g. for logic elements and high-density non-volatile magnetic random access memory[1, 2, 3, 4]. Like conventional semiconductors, DMSs have a band gap between the valence band and the conduction band of 0.4 - 3.5 eV, but by adding a low concentration of magnetic ions to the undoped parent semiconductor, it is possible to make them ferromagnetic. As possible parent materials, the III-V semiconductors such as GaAs are already in use in a variety of electronic and optoelectronic devices, including cellular phones (transistors), compact disk drives (semiconductor lasers), and many other applications[5]. By adding the electron spin as an additional degree of freedom III-V–based DMSs are being used to explore a new field in semiconductor physics and technology, where both semiconducting and magnetic properties play critical roles[1,2,3,4].

The most investigated DMS to date is probably $Ga_{1-x}Mn_xAs$[1, 2]. The role played by the Mn dopant can be explained by the Zener mechanisms[6, 7, 8], which are a combination of double exchange and *p-d* exchange interactions. In the double-exchange mechanism, Mn-impurity states form within the band gap, close to the valence band maximum (VBM); in the *p-d* exchange mechanism, Mn hybridizes with GaAs valence band, and induces hole doping and magnetism. Recent theoretical[3] and experimental[9, 10, 11] studies indicate that both mechanisms contribute to the emergence of ferromagnetism in $Ga_{1-x}Mn_xAs$, including hard x-ray angle-resolved photoemission (HARPES) measurements[9] and soft x-ray resonant ARPES[10,11].

To fully understand the electronic and ferromagnetic properties of DMSs, a thorough characterization of their electronic structure is essential. The most direct method for obtaining this information is photoemission spectroscopy. However, when the bulk properties of the materials are significantly different from their surface properties, the extremely low probing depth of conventional vacuum ultraviolet or soft X-ray photoemission can be a disadvantage. An increase of the photoelectron escape depth, and therefore higher bulk sensitivity, can be achieved by using core-level hard-X-ray photoemission spectroscopy (HXPS, HAXPES)[12, 13] and HARPES[9, 14].

In this paper, we present a study of the electronic structure of $Ga_{1-x}Mn_xP$ and GaP samples by HXPS and HARPES, in combination with fully relativistic one-step model photoemission calculations. Our results allow us to demonstrate in detail the influence of the Mn dopant in this DMS sample. By HARPES we directly observed the emergence of Mn-induced states over the full valence-band region, confirming the combination of the double-exchange and *p-d* exchange mechanisms as the origin of the ferromagnetism in this material. Satellite and multiplet structures in Mn 2*p* and 3*s* core-level spectra also indicate a very similar bonding site (i.e. substitutional for Ga) and spin state to that seen in (Ga,Mn)As.

## II.    Experimental Procedures

The reference sample, gallium phosphide (GaP) is a semiconductor with an indirect band gap of $\Delta E_g = 2.26$ eV. GaP has a zinc blende crystal structure with a lattice constant $a = 5.451$ Å[15]. Our n-type sample had (001) surface orientation, and about $1 \times 10^{16} - 10^{17}$ cm$^{-3}$ concentration.

The $Ga_{1-x}Mn_xP$ DMS samples were grown on an n-type (001) GaP crystal equivalent to the reference sample using the ion implantation and pulsed laser melting (II-PLM) method[16]. The $x$ in these samples is defined as the peak substitutional manganese fraction, occurring between 20 and 30 nm below the surface, as determined by secondary ion mass spectroscopy (SIMS) analysis. In this paper, we primarily studied a doped $Ga_{1-x}Mn_xP$ sample with the peak Mn concentration near $x = 0.02$. Data was also obtained for other Mn concentrations, $x$ values of 0.032 and 0.041[17]. The Curie temperatures of these samples were: 18 K for x = 0.020, 52 K for x = 0.032, and 60 K for x = 0.041. Quantitative analysis of Ga, Mn, and P core-level HXPS intensities, including variable-angle HXPS, was used to check these compositions, and to derive the depth profile of Mn concentration in our samples[17].

Transmission Electron Microscopy (TEM) analysis has shown that II-PLM $Ga_{0.98}Mn_{0.02}P$ is single crystalline, though in the as grown state the first ~25 nm from the surface are highly defective. This poorly regrown layer is readily removed by etching in concentrated HCl for ~24 hours[18], resulting in epitaxial $Ga_{0.98}Mn_{0.02}P$ grown on top of GaP, and the Mn distribution is approximately 100 nm in width[19]. Our variable-angle HXPS results are qualitatively consistent with this and show that, after a few cycles of HCl etching, the Mn concentration was very low at the surface, rising to approximately 5% at a depth of about 8 Å and then decreasing linearly to zero at a depth of approximately 500 Å[17]. In addition, Rutherford Back Scattering (RBS) and particle-induced X-ray emission (PIXE) suggest that 70%−85% of the Mn atoms occupy substitutional Ga sites and that the remaining Mn atoms are at random locations within the lattice[20,21]. Our Mn core level spectra, to be discussed below, also strongly point to only one type of Mn bonding site.

We used HXPS and HARPES techniques to determine the composition and the electronic structure of these samples. HXPS allows us to study bulk-like electronic properties of these samples. Conventional XPS is generally a surface sensitive technique; however, with the use of hard X-ray photon energies in the few keV regimes, more bulk-like information can be obtained. The electron inelastic mean free path (IMFP), obtained using X-ray optical data[22], gives an estimation of the average depth from which the elastically scattered electrons are emitted from the sample. Due to the low IMFPs of the photo emitted electrons in conventional ARPES[23] , which are 10 Å or lower, spectra can be strongly influenced by surface effects[14]. In order to avoid this problem and to study the bulk-like electronic properties of the DMS, we have performed ARPES at hard X-ray energies around 3 keV. At these photon energies the IMFP for GaP valence electrons is predicted by the TTP-2M[22] method to be about 53 Å. Since GaP has a lattice constant of $a =$ 5.415 Å, this photon energy should yield a probing depth of about 9-10 unit cells, with an effective overall sensing depth of about three times that.

HXPS and HARPES measurements were carried out at the Advanced Light Source (ALS) at the Lawrence Berkeley National Laboratory (LBNL), Berkeley, California, USA[24] and at SOLEIL Synchrotron in St. Aubin, France[25].

At the ALS, the data were collected at Beamline 9.3.1 using the Multi-Technique Spectrometer/Diffractometer end station[26]. The end station is equipped with a Scienta SES 2002 analyzer and operates with a custom-built five-axis variable-temperature sample goniometer, allowing movement in *x, y, z,* and the polar and azimuthal angles $\theta$ and $\phi$, and cooling of the sample with liquid nitrogen (LN) to about 80 K. In angle-resolving mode, the spectrometer spans ~$\pm 6°$. All HXPS and HARPES measurements at this beamline were performed at 2905 eV and 2550 eV. Total energy resolution was set at approximately 300 meV.

At SOLEIL, the data were collected at beamline GALAXIES using a Scienta R4000 spectrometer and a four-axis manipulator connected to a cryogenic cooling system[27]. The HARPES data were collected with the sample cooled to about 30 K. The angle-averaged spectra for GaP and $Ga_{0.98}Mn_{0.02}P$ samples was measured at photon energies of 2905 eV, 3100 eV and 3270 eV. This angle-resolving mode in this spectrometer spans ~±25°. Total energy resolution for the HARPES measurement was set at approximately 200 meV.

Some preliminary HXPS data for doped and undoped GaP was also obtained at the revolver undulator beamline, BL15XU[28]. of SPring-8. The end station is equipped with a VG Scienta R4000 analyzer as described in Ref. 24. The photon energy was set at 5950 eV for the angle-integrated HXPS measurements, and total energy resolution was set to 240 meV at 30 K. For preliminary HARPES data were taken at 5950 eV at 30 K. Total energy resolution for HARPES was set to 300 meV.

## III.    Theoretical Calculations

### A.    Free-electron final-state calculations (FEFS)

We have used free-electron final state (FEFS) calculations to obtain the trajectory in $k$-space that is probed in the HARPES measurements. This model has been described elsewhere[47], and has been used successfully as a first stage of HARPES analysis in prior studies. The initial-state band structure of GaP was calculated using the linearized augmented plane-wave (LAPW) WIEN2k all-electron ab-initio package[29] at a lattice constant of 5.451 Å and with the PBE-GGA exchange correlation potential[30], and compared with varying experimental geometries as to slight tilts away from the expected ideal setup geometry (see inset in Figure 1) to the experimental HARPES maps. This procedure allowed tuning to determining the experimental geometry, including possible small tilts of the sample away from high-symmetry directions along three orthogonal axes. The photon energy was fixed at 2905 eV and the inner potential at 10 eV. Non-dipole effects due to the photon momentum were also included.

### B.    One-step photoemission

In addition to the previously mentioned LAPW calculations, we used a one-step time-reversed-LEED-state calculation of the photoemission, based on the full-relativistic multiple scattering Korringa-Kohn-Rostoker (KKR) Green's function method, as implemented in the spin-polarized relativistic SPRKKR code[31, 32]. In addition to the LAPW method, the coherent potential approximation (CPA) was introduced in order to describe Mn doping at random substitutional sites in the lattice. This approach permits describing chemical disorder in the system in a much simpler way than laborious super cell approaches. As a first step of our investigations, we performed ground state self-consistent calculations for the GaP and $Ga_{0.98}Mn_{0.02}P$ systems. All calculation parameters, such as for example the numbers of k-points, and the exchange correlation functional, have been taken to be as close as possible to those used in the LAPW calculations. For the angular momentum expansion of the Green's function, a cutoff of $l_{max} = 3$ was applied. Both methods gave essentially identical results for ordered GaP with respect to band structure and density of states (DOS). The spectroscopic analysis is based on the fully relativistic one-step model of photoemission in its spin-density matrix formalism, as embodied in the SPRKKR code. The photoemission calculation itself is based on multiple-scattering theory, using explicitly the time-reversed low-energy electron diffraction (LEED) method to calculate the initial and final states for a semi-infinite atomic half-space.

These one-step photoemission calculations include all matrix element effects, all multiple scattering effects in the initial and final states, the effect of the photon momentum vector, and the escape depth of the photoelectrons via an imaginary part in the potential function. For example, we had to include about 70 layers of $Ga_{0.98}Mn_{0.02}P$ (001) surface to be able to converge the photocurrent to $10^{-5}$ relative to the surface for the photon energy of 2.9 keV, using an imaginary part of the inner potential of $V_{0i} \approx 8.0$ eV, which permits converging to the bulk electronic structure in depth. Further details concerning the CPA implementation of the one-step model for HARPES are presented elsewhere[33, 34, 35]. The experimental geometry for the calculations was taken from the fit of FEFS calculations to experiment. A small tilt of the sample and azimuthal rotation, more noticeable for the undoped GaP, were found using the FEFS model.

## IV. Experimental results and comparison to theories

### A. Survey scans and Ga, P and Mn core levels

Figure 1 shows the survey spectra of GaP and $Ga_{0.98}Mn_{0.02}P$ collected at the photon energy 2905 eV. All major features of the spectra can be assigned to core level signals that were expected from the composition of the samples. The C $1s$, O $1s$, and very weak N $1s$ signals indicate surface contamination probably due to exposure to ambient air during transportation and the transfer process to the vacuum chamber after the HCl etch. Using the SESSA[36] surface analytical program, we were able to determine that the contaminant layer at the surface is about 10 Å thick. For the doped sample, a very weak Cu $3p$ doublet signal at ca. 935 eV and 955 eV was noticed, and confirmed, again making use of SESSA, to be a strictly surface contaminant layer about 0.2 Å thick; this should not affect any of our conclusions from these more bulk sensitive HXPS and HARPES measurements. The inset in Figure 1 shows the experimental geometry for our HXPS and HARPES measurements at the ALS.

Figure 2 shows the detailed core level spectra of Ga 2$p$ and P 2$p$ for GaP and Ga$_{0.98}$Mn$_{0.02}$P. These two core levels exhibit the expected spin-orbit splitting of $\Delta E = 26.71$ eV and $\Delta E = 0.78$ eV, respectively. These two core-levels are similar for the two samples in shape, intensity, and in binding energy (E$_B$). Further investigation[17] of the Ga 2$p$ and P 2$p$ peaks of other DMS samples with $x$ ranging from 0.020 up to 0.042, show that the Ga 2$p$ intensities varies as a function of the Mn concentration according to $1$-$x$, (% doping), using the P 2$p$ intensities as a constant reference for normalization. These results are consistent with prior RBS results indicating that the Mn atoms predominantly occupy substitutional sites of Ga[18]. In Figure 2 we also note a strong inelastic loss feature at ca. 1134 eV due to the well-known plasmon excitation of GaP at 16.6 eV[37].

Since Mn is expected to be the element introducing magnetic properties to the material, it is important to look in more detail at the core levels of Mn, as shown in Figures. 3 and 4. Figure 3 shows the Mn 2$p$ core-level spectra for the doped samples measured at ALS and SPring-8, for different Mn dopant concentration, and at different photon energies. For Ga$_{0.98}$Mn$_{0.02}$P, measured at ALS at 2905eV photon energy (Figure 3(a)), the two main peaks, at E$_B$ of 650.61 eV and 638.52 eV, are the 2$p_{3/2}$ and the 2$p_{1/2}$ states with a spin-orbit splitting of $\Delta E = 12.09$ eV. The Mn 2$p_{3/2}$ region further shows an extra feature, at E$_B$ of 639.56 eV that has been interpreted as a well-screened final state, for which the screening electrons are highly delocalized and near the Fermi energy[38, 39]. This extra feature has been seen not only in the DMS material Ga$_{1-x}$Mn$_x$As[39], but in a number of other materials, including ferromagnet/superconductor heterostructures[40]. A broader satellite peak, forms near the main peaks at approximately 643.2 eV and 656.7 eV. These satellites are due to well-screened core-hole states[39]. The component shown in green at far left no doubt has some contribution from the GaP plasmon loss at 16.6 eV, as well as the expected well-screened states. Noteworthy here is that these spectra are to within experimental accuracy identical, with only a possible difference in the relative intensity of the sharp highly screened peak for the 5950 eV results; this can be due to the deeper probing at this energy and also better energy resolution, as it is known that such features are only clearly visible with higher-energy excitation, due to their delocalized origin[35].

Figure 4 displays the Mn $3s$ photoemission spectra for the same samples. Although somewhat difficult to resolve due to overlap at higher $E_B$ with the tail of the Ga $3p$ peak, the Mn $3s$ core level shows a well-known doublet due to multiplet splitting[41], with a value of $\Delta E_{3s} = 5.40$ eV measured at SPring-8 (Figure 4(a)), which agrees with prior studies of various Mn compounds[42] and corresponds to a Mn valence of about $3d^5$, as expected for this dopant. Similar results from the other investigated samples with different Mn concentrations measured at ALS (Figures 4(b)-(d)), indicate that the multiplet splitting exchange energy $\Delta E_{3s}$ is independent of the amount of the dopant, yielding values of 5.79 eV, 5.78 eV, and 5.72 eV for 2.0%, 3.2%, and 4.1%, respectively[17]. These are slightly different from the value obtained with 5950 eV excitation, and are mostly due to the greater difficulty of fitting, but possibly also to slight surface alteration of the Mn state due to the lower excitation energies. Consistent with the Mn $2p$ spectra, the Mn $3s$ spectra also indicate a single substitional site for the dopant atoms, with charge and spin state essentially identical and constant with these low dopant levels.

### B.     HARPES at 2.9 keV and comparison to theories

As noted above, the HARPES measurements at the ALS were performed at the same photon energy as the HXPS data ($h\nu = 2905$ eV) shown above, for which the probing depth is about 9-10 unit cells. Sample cooling to about 80 K is especially critical at such high photon energies, since the dispersive direct-transition features of the band structures are smeared out and suppressed by phonon-assisted non-direct transitions. The phonon smearing can be estimated using a photoemission Debye-Waller factor[9, 14], which predicts the fraction of the direct transitions in a solid for a given photon energy and temperature. The Debye-Waller factor is calculated from the formula:

$$W(T) = \exp[-(\frac{1}{3})g_{hkl}^2 < U^2(T) >]\,, \qquad [1]$$

where $g_{hkl}$ is the magnitude of the bulk reciprocal lattice vector involved in the direct transitions for a given photon energy, and $<U^2(T)>$ is the three-dimensional mean-squared vibrational displacement for a given temperature[43]. At high photon energies and high temperatures[9,14], the effects of the phonon smearing hinder the observation of dispersive features in the angle–resolved photoemission signal, and instead the spectra are dominated by the matrix-element weighted density of states (MEWDOS). However, by cooling the sample to about 80 K, we increase the fraction of the direct transition in the solid to about 25% (Debye-Waller factor $\approx 0.25$) for GaP at 2905 eV, which results in a clearer definition of the dispersive bands on top of the DOS-like background. Such Debye-Waller estimates have also been found for some cases to be too conservative, with the effective number of direct transitions being higher than expected[44]. This enabled our study of the bulk band structure of GaP and $Ga_{0.98}Mn_{0.02}P$.

As a first step toward understanding the experimental HARPES data, we describe the calculations based on a simple but useful model of rigorous **k**-conserving or direct transitions from an initial ground state band structure to a strictly free-electron final state (FEFS), as used successfully in several prior studies [9,14,45]. That is, the band structure provides band energies $E_i(\boldsymbol{k_i})$, which are coupled to free-electron final states at $E_f(\mathbf{k_f})$ by the following two conservation formulas:

$$E_f(\boldsymbol{k}_f) = E_i(\boldsymbol{k}_i) + h\nu = \frac{\hbar^2 k_f^2}{2m_e} - V_0 + \phi_s = \frac{\hbar^2 \boldsymbol{K}^2}{2m_e} + \phi_s \qquad [2]$$

and

$$\boldsymbol{k}_f = \boldsymbol{k}_i + \boldsymbol{k}_{h\nu} + \boldsymbol{g}_{hkl} \qquad [3]$$

where $E_f(\boldsymbol{k}_f)$ is the final electronic kinetic energy, $E_i(\boldsymbol{k}_i)$ is the initial energy relative to the Fermi energy, $h\nu$ is the photon energy, $m_e$ is mass of the electron, $V_0$ is the inner potential, $\phi_s$ is the work function, $\boldsymbol{k}_i$ is the initial-state wave vector in the reduced Brillouin zone (BZ), $\boldsymbol{k}_f$ is the final-state wave vector inside the crystal, $\boldsymbol{K}$ is the wave vector of the photo emitted electron in vacuum, $\boldsymbol{k}_{h\nu}$ is the wave vector of the photon, and $\boldsymbol{g}_{hkl}$ is the bulk reciprocal lattice vector.

Figure 5 shows the schematic of the reciprocal space geometry for our HARPES experiment and includes the solid angle subtended by the detector slit. An extended BZ diagram is presented as the background scale. For an inner potential of $V_0 = 10$ eV, a work function of $\phi_s = 5\ eV$ and our experimental photon energy of $h\nu = 2905\ eV$, the magnitude of the final-state photoelectron wave vector inside the crystal for an electron at the Fermi level is[12,44]:

$$k_f (\mathring{A}^{-1}) = 1/\hbar \sqrt{2m_e(h\nu + V_0 - \phi_s - E_b)} = 27.62\,\mathring{A}^{-1} = 11.98(\frac{4\pi}{a}) \qquad [4]$$

The magnitude of the photon momentum can be calculated as[44]:

$$k_{h\nu}(\mathring{A}^{-1}) = 0.000507 h\nu(eV) = 1.473\ \mathring{A}^{-1} = 0.639(\frac{4\pi}{a}). \qquad [5]$$

The angle between photon incidence and electron emission is $90°$, and the incident angle is around $0.3°$; thus the average emission direction is very near to the [001] normal of the sample (cf. inset of Figure 1). The distance $\Gamma - K - X - K - \Gamma$ in the extended-zone scheme along the direction of the detector slit is approximately $6.8°$, and has been drawn to reflect this. The quantity $\Delta$ is the shift in $\boldsymbol{k}$-space caused by the photon momentum, about $3.08°$.

Figures 6(a),(b) show the raw data HARPES experimental spectra of (a) GaP and (b) $Ga_{0.98}Mn_{0.02}P$, measured in the real-space geometry shown in Figure 1, and thus also the reciprocal space geometry shown in Figure 5, but with a slightly different tilt for each sample, which we will discuss later. The data represent single detector images spanning approximately 12°, and they clearly show dispersive features that are overlaid with flat bands of intensity that are DOS-like that are also expected to show x-ray photoelectron diffraction (XPD). The HARPES images shown in Figures 6(c),(d) are obtained from the raw data after a two-step correction process, used previously for HARPES data[9,14 , 46]. First we normalized by dividing the image by the angle average at each detector energy, which approximately resembles a division by the MEWDOS; these normalization curves are shown at right in Figures 6(a),(b) in orange. Second we normalized by dividing by the energy average at each detector angle, to correct for intensity modulations which are dominated by XPD; these curves are shown at the bottom of Figures 6(a),(b), in blue. Such normalizations remove the constant-angle (MEWDOS) and constant-energy (XPD) modulations from the detector image and highlight the dispersive bands, which are now very clearly shown in Figures 6(c),(d).

In the corrected HARPES experimental results we notice significant differences in the band dispersion between the two samples. Partially these differences are due to a slight misalignment of the samples, to which HARPES data are extremely sensitive. Since the $\boldsymbol{k}_f$-vector has a length of ~12 BZs, even a tilt of the surface normal in the order of 0.1° results in a sizable momentum offset within the reduced zone scheme (cf. Figure 5). To determine the exact experimental geometry of each sample, we used the FEFS theory in combination with density functional theory (DFT) calculations[47]. Using the DFT program WIEN2K[29], we calculated the bulk band structure of GaP. The combination of DFT with the FEFS model then allows us to calculate the expected band dispersion for the HARPES experiment, which strongly depends on the sample alignment.

By varying the geometry for the FEFS calculations, using a very fast algorithm that will soon be published[47], we were able to match the band dispersions that were obtained by the FEFS calculations to our experimental data, and determine the sample tilt, a rotation around [110] as shown in Figure 1, within an accuracy of about $0.1°$.

Figures 7(a),(b) now show the HARPES experimental data of (a) GaP and (b) $Ga_{0.98}Mn_{0.02}P$ together with the corresponding calculations from FEFS theory, with the experimental geometry optimized to fit the experimental dispersions. There is excellent agreement between the FEFS results and the main experimental features, with some disagreement at the edges of the images that are due to detector non-linearities. Figures 7(c)-(e) further show the paths within $\boldsymbol{k}$-space for the HARPES experiment according to the FEFS model for the optimized geometries from three different perspectives. For the $Ga_{0.98}Mn_{0.02}P$ sample we find within the error of our determination that the entrance slit of the analyzer is very precisely aligned along the [110] direction; that is there is no tilt or azimuthal rotation. For the GaP sample, however, we had to introduce a tilt of the sample normal by $0.6°$, as well as an azimuthal rotation by $2°$ to find the best agreement. Therefore, the path in k-space for GaP differs slightly from the $\Gamma - K - X - K - \Gamma$ direction, as shown in Figures 7(c)-(e).

Once the optimized geometries are obtained using the FEFS model, we have proceeded to simulate the HARPES spectra by using one-step photoemission theory, as introduced above[34]. In addition to the corrected HARPES experimental results of GaP (Figure 6(c)) and of $Ga_{0.98}Mn_{0.02}P$ (Figure 6(d)), Figures 6(e),(f) show the results of our fully relativistic one-step photoemission calculations. Unlike the FEFS calculations, the one-step calculations incorporate Mn doping as well as surface and matrix element effects, and therefore allow for a thorough, quantitative comparison to the experimental HARPES data.

The agreement between the experimental data and the theoretical calculations in Figure 6 is remarkable for both samples. Figure 7 also shows that the band dispersions observed in the one-step calculations is also in good agreement with FEFS calculations. The experimental data and theories also show evident differences between GaP and $Ga_{0.98}Mn_{0.02}P$, with the major one in both the experimental data and the one-step calculations being the change in the sharpness of the bands. The $Ga_{0.98}Mn_{0.02}P$ bands are broadened due to the presence of Mn impurities that disturb the long-range translational order of the host GaP crystal. This sort of difference has been noted in a previous HARPES study of GaAs and $Ga_{0.97}Mn_{0.03}As$[9].

### C. Angle-integrated data and matrix-element weighted densities of states

As another way of looking at such HARPES results, we show in Figure 8(a) the angle-integrated valence band spectra of GaP and $Ga_{0.98}Mn_{0.02}P$ as obtained from the raw data of Figures 6(a),(b) before the two step normalization process, and their spectral difference. Although these spectra were measured using the angular mode of the electron analyzer, the D-W factor of about 25% at such high photon energies suggests that they should represent a good approximation of the MEWDOS. Figure 8(b) shows the corresponding theoretical spectra and their differences obtained from one-step CPA calculations. All of these spectra were normalized by the mean of the total area of the valence-band. The experimental and the theoretical spectra are in good agreement in terms of all major features. Figures 8(c),(d) show a direct comparison of the difference plots ($Ga_{0.98}Mn_{0.02}P$ - GaP) from Figures 8(a),(b). Again, there is a good agreement between experiment and theory for these difference plots, with the features labelled 1 - 5 showing theoretical counterparts 1' - 5'. These results for differences are also in general agreement with those of Gray et al. for ($Ga_{0.97}Mn_{0.03}As$ − GaAs)[9].

These difference plots clearly indicate that the presence of Mn introduces changes in the electronic structure over the full valence-band energy range, including small shape changes and peak shifts. The appearance of feature 1 in the experimental data and its counterpart 1' in theoretical calculations, above the valence band maximum clearly show the presence of Mn-induced impurity states.

In Figure 9, we explore the calculations in more detail, with total and orbital-projected DOSs for $Ga_{0.98}Mn_{0.02}P$ calculated using the CPA model in the local density approximation (LDA) framework, as was done for the one-step photoemission[48, 9]. The results are shown in Figures 9(a),(b) for the total DOS for both spin-down and spin-up states, as well for Ga, P, and Mn, and in Figures 9(c),(d) in more detail for the *s, p,* and *d* DOS for Mn. These results show that Mn *3d* spin-up in the ferromagnetic state contributes significantly to the electronic structure at the Fermi level, but also contributes significantly to the entire valence-band region, down to the pseudo bandgap from 7 to 9.5 eV.

To further illustrate the applicability of the MEWDOS picture, we consider the angle-integrated difference curves from another set of HARPES results for ($Ga_{0.98}Mn_{0.02}P − GaP$) obtained at the Soleil Synchrotron with a detector spanning a much wider range in angle of ~40°. The spectra were measured at three different photon energies and thus average $k_f$ values, 2905 eV (matching the ALS data) with $\boldsymbol{k}_f$ = 27.59 Å⁻¹, 3100 eV with $\boldsymbol{k}_f$ =28.50 Å⁻¹, and 3270 eV with $\boldsymbol{k}_f$ =29.27 Å⁻¹. The latter two thus correspond to a translation of the $\boldsymbol{k}$-path by about 0.91 Å⁻¹ and 1.69 Å⁻¹ along [001], respectively, or about a 39% and 73% shift of the BZ dimension 2.31 Å⁻¹ along that direction. The same normalization was used for the experimental angle-integrated valence-band data and for the spectral differences presented in Figure 8.  In Figure 10 a global comparison of the angle-integrated valence–band data is shown, including all the experimental data for the differences ($Ga_{0.98}Mn_{0.02}P - GaP$), the one-step theory calculations, and the difference from the prior study of ($Ga_{0.97}Mn_{0.03}As - GaAs$) at energy 3240 eV[9]. The data from Soleil in Figures 10(d)-(f) are remarkably similar, and almost directly superimposable, even though the ***k***-paths are shifted significantly along [001]; they also do not show feature 5 clearly, and this may be due to the more complete BZ averaging involved, compared to the ALS results. Overall these observations give further support to the interpretation of our data as a good approximation of the MEWDOS, due to the high degree of phonon scattering as represented by the D-W factors, as noted above. The Soleil data are also in qualitative agreement with our results from the ALS (Figure 10(a)), and in even better agreement with theory (Figure 10(b)), probably due to the larger angular acceptance of the analyzer, which samples a larger fraction of the reciprocal space. That feature 1 is more pronounced in the $Ga_{0.97}Mn_{0.03}As - GaAs$ difference[9] may indicate a fundamental difference in the effects of Mn in the GaP-based material. In Figure 10(c), we also note that the prior experimental differences of ($Ga_{0.97}Mn_{0.03}As - GaAs$) are very similar, although not identical, to those of ($Ga_{0.98}Mn_{0.02}P - GaP$) reported in this study. This suggests that (Ga,Mn)P behaves very similarly, although not identically, to (Ga,Mn)As, as might be expected based on the similar ionic radii and electronegativies of P and As, and more detailed calculations of the energy differences between the ferromagnetic and paramagnetic states for these DMS materials at dopant levels similar to ours (cf. Figure 12 in Ref. 3).

## V.    Conclusions

We have used core-level HXPS and valence HARPES in combination with free-electron final state and one-step photoemission theory to study the composition and electronic structure of $Ga_{0.98}Mn_{0.02}P$ and GaP. The Mn $2p$ core spectrum in particular exhibits a well-screened final state which has been seen in prior studies. Mn $3s$ also shows a doublet due to multiplet splitting, indicative of the same magnetic moment as found in (Ga,Mn)As. The near identity of both of these sets of spectra, including three different compositions, indicates a dominant single substitutional site for the Mn.

We have also confirmed that angle–resolved photoemission with high energy X-rays allows for detailed bulk-like electronic structure determination. We have applied HARPES to doped $Ga_{0.98}Mn_{0.02}P$ and GaP samples. Differences were observed between the electronic structures of GaP and $Ga_{0.98}Mn_{0.02}P$, including smearing of the bands due to the presence of Mn impurities that disturb the long-range translational order of the host GaP crystal. Changes throughout the entire valence band were present, including the appearance of a Mn-induced impurity band, at approximately 0.44 eV below the Fermi level. Free-electron final-state calculations were used to determine the experimental geometries of each sample. Then, experimental data and one-step photoemission calculations were compared to each other and are in very good agreement in terms of the spectral differences between $Ga_{0.98}Mn_{0.02}P$ and GaP. Angle-integrated HARPES results at several energies agree very well with one-step photoemission theory, particularly when a wider detector angle of about 40° is integrated over.

These HARPES results for $Ga_{0.98}Mn_{0.02}P$ thus represent a continuation study of the group III-V DMS materials[9] and shows good agreement with the most studied DMS material: $Ga_{1-x}Mn_xAs$. The results presented here add to our understanding of these DMS materials and the significance of the Mn dopant in them. Applying such HARPES measurements to other semiconductors doped with Mn, e.g. $Ga_{1-x}Mn_xN$, or other complex multi-component materials, should assist in further elucidating their electronic structures.

**Acknowledgments**

C.S.F. has been supported for salary by the Director, Office of Science, Office of Basic Energy Sciences (BSE), Materials Sciences and Engineering (MSE) Division, of the U.S. Department of Energy under Contract No. DE-AC02-05CH11231, through the Laboratory Directed Research and Development Program of Lawrence Berkeley National Laboratory, through a DOE BES MSE grant at the University of California Davis from the X-ray Scattering Program under Contract DE-SC0014697 (for experiments at the Advanced Light Source and for travel support to carry out experiments at Soleil), and through the APTCOM Project, "Laboratoire d'Excellence Physics Atom Light Matter" (LabEx PALM) overseen by the French National Research Agency (ANR) as part of the "Investissements d'Avenir" program. J.M. would like to thank CEDAMNF project financed by Ministry of Education, Youth and Sports of Czech Rep., project No. CZ.02.1.01/0.0/0.0/15.003/0000358. The HXPS measurements at SPring-8 were performed under the approval of NIMS Synchrotron X-ray Station (Proposal Nos. 2009A4906, 2010A4902, and 2010B4800), and were partially supported by the Ministry of Education, Culture, Sports, Science and Technology (MEXT), Japan. Graduate student researcher-work study and departmental fellowship was awarded for A.K at UC Davis. A.R. was funded by the Royal Thai Government and C.C. was funded by GAANN program through UC Davis Physics Department.

**Figures**

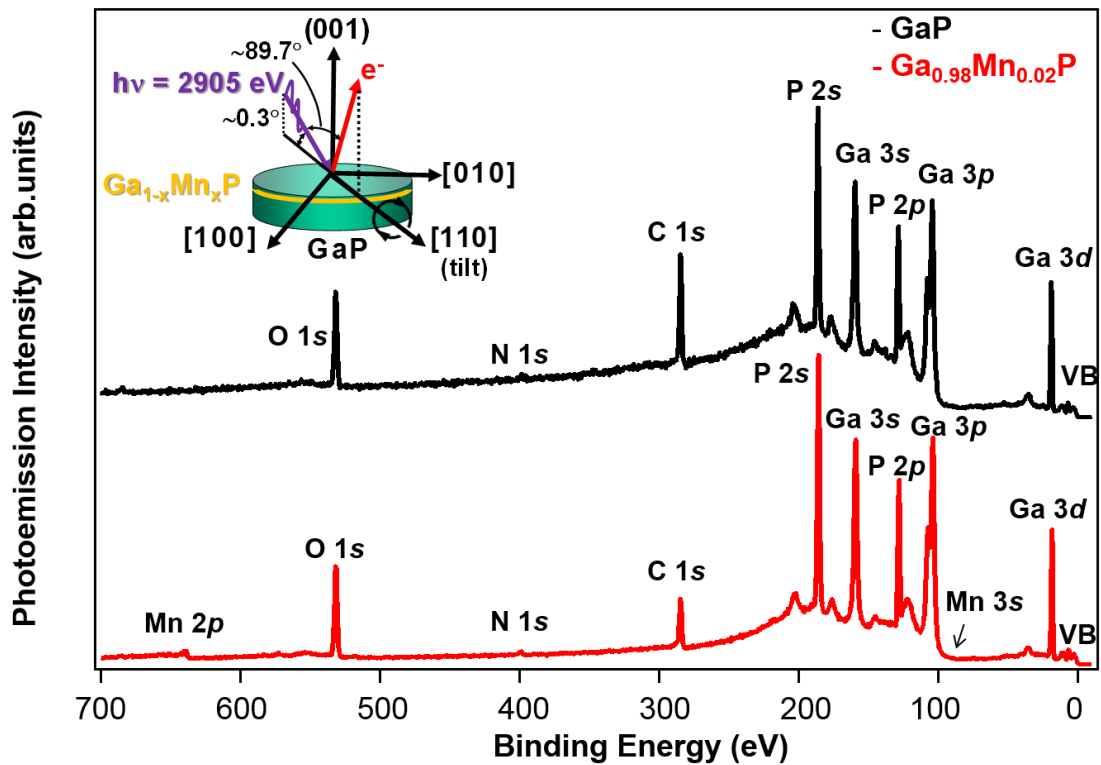

FIG. 1. Survey spectra of GaP and Ga$_{0.98}$Mn$_{0.02}$P at a photon energy of 2905 eV. O, C and N are contaminations. The inset shows the experimental geometry for these experiments.

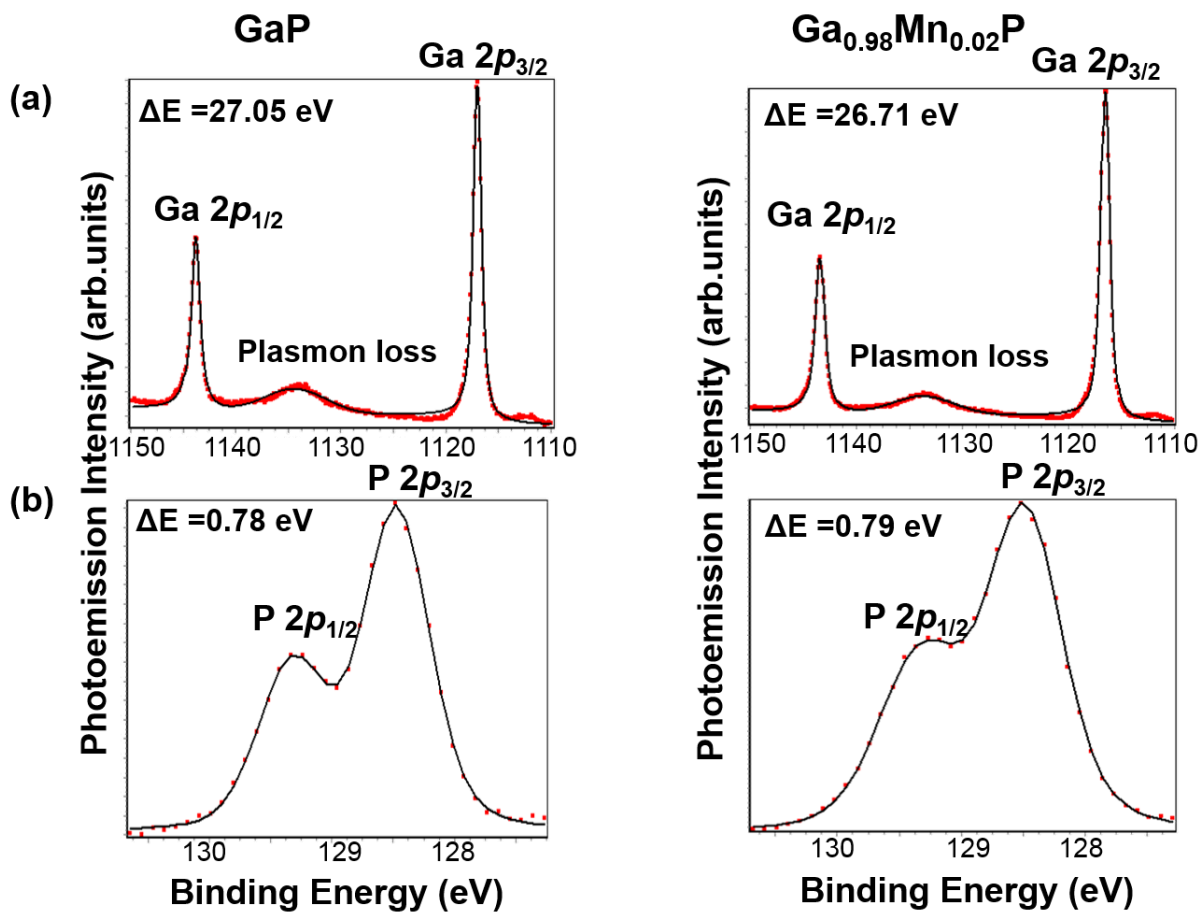

FIG. 2. **(a)** Core-level spectra of Ga 2*p* and **(b)** core-level of P 2*p* in GaP and Ga$_{0.98}$Mn$_{0.02}$P samples, respectively. Red curves experimental data and black curves fitted core-levels.

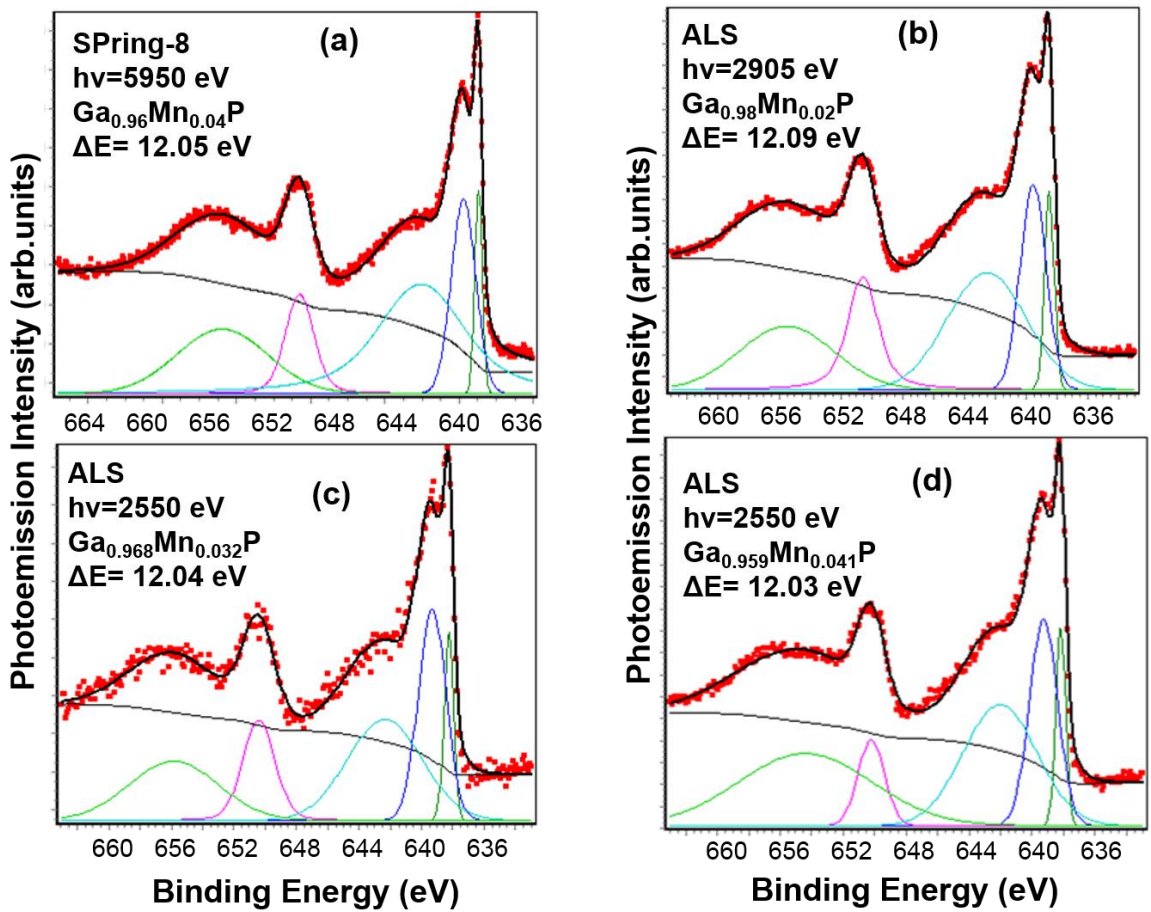

FIG. 3. Core-level spectra of Mn $2p$ from the (Ga,Mn)P samples, for different Mn % dopant and at different photon energies, from **(a)** SPring-8 at 5950 eV and **(b)-(d)** ALS at 2905 eV, 2550 eV, and 2550 eV, respectively. In **(b)-(d)**, the Mn concentration also varies over $x = 0.020$, 0.032, and 0.041. Satellite features are shown near the main peaks, including a sharp screening feature at lowest binding energy. The set of peaks used to fit these spectra are also shown.

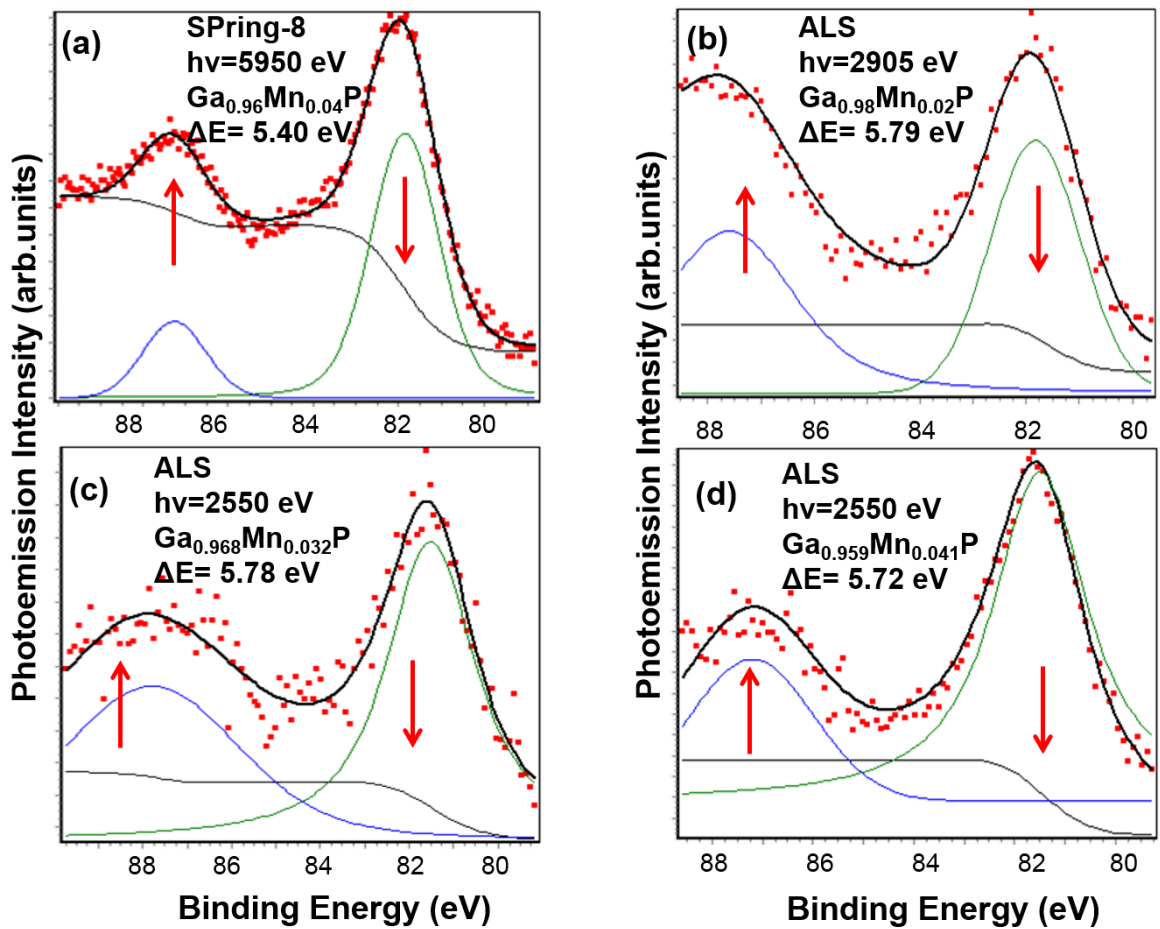

FIG. 4. As FIG. 3, but for Mn 3*s* spectra. The dominant direction of the spin of the photoelectron with respect to the 3*d* moment on Mn is indicated. The set of peaks used to fit these spectra are also shown.

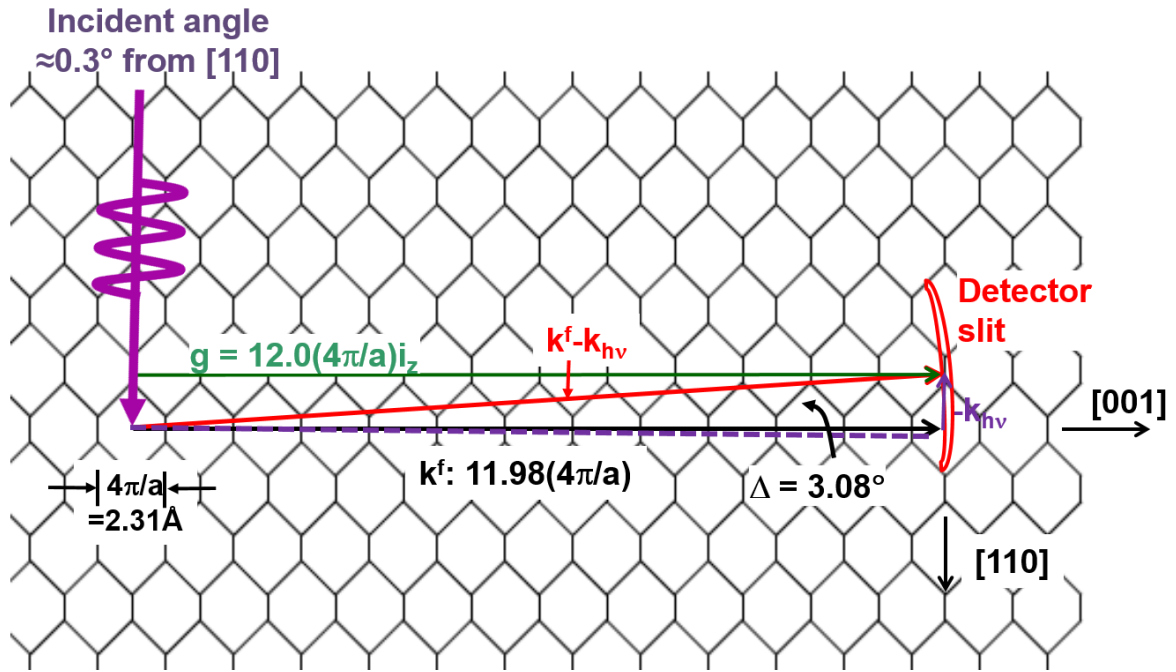

FIG. 5. The experimental geometry of the HARPES measurement in an extended zone picture, with various key vectors and angles shown.

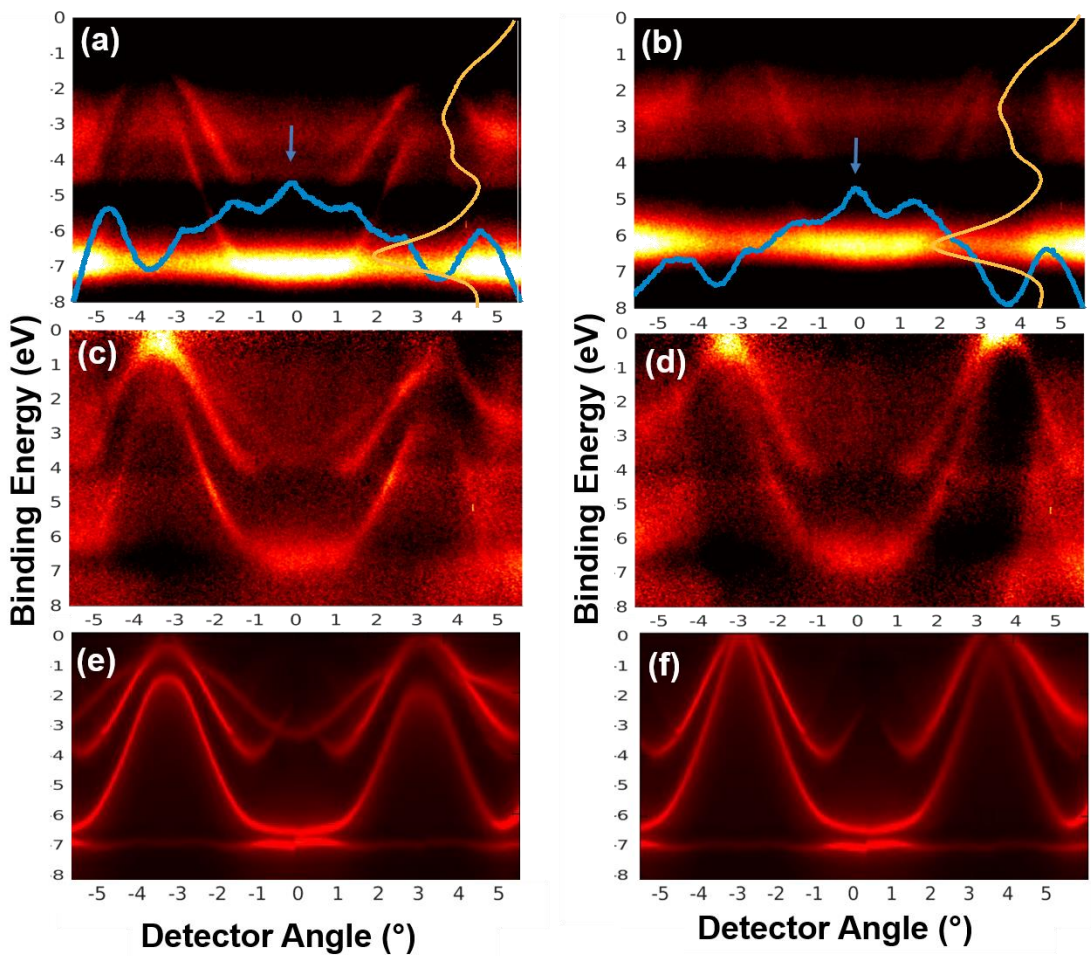

FIG. 6. **(a)** Raw HARPES experimental measurements from GaP(001) **(b)** $Ga_{0.98}Mn_{0.02}P$, with the angle-averaged MEWDOS-like correction curve shown in <span style="color:orange">orange</span> at right, and the energy-averaged XPD correction curve in light <span style="color:blue">blue</span> at the bottom, and the arrows indicating the [001] direction associated with a forward scattering peak in the <span style="color:blue">blue</span> XPD curve, **(c), (d)** data from **(a)** and **(b)** after division by the two correction curves so as to remove the DOS and XPD effects. **(e),(f)** One-step photoemission calculations of the HARPES spectra, carried out along the path in k-space derived from FEFS calculations (see next figure).

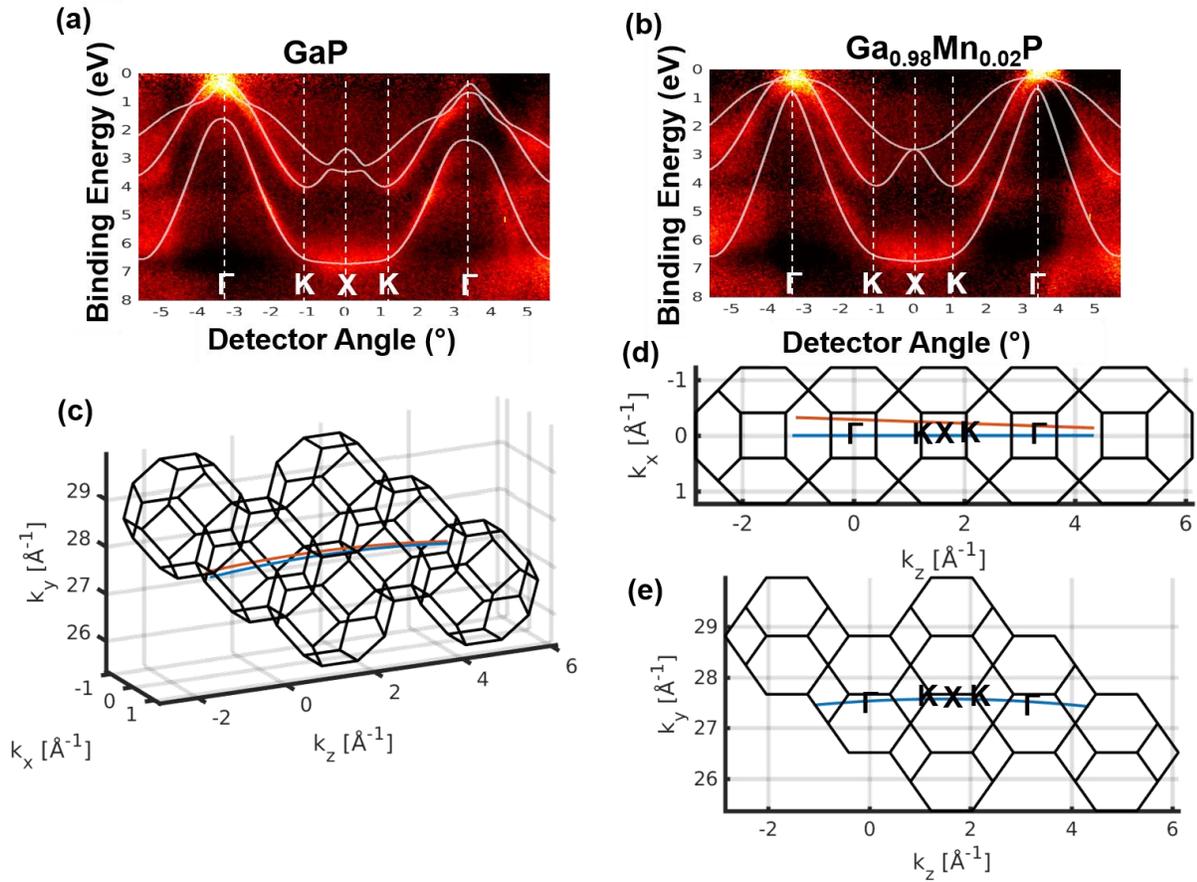

FIG. 7. **(a),(b)** Curves that are the results of the free-electron final-state calculations are superposed on the corrected experimental HARPES data for GaP and $Ga_{0.98}Mn_{0.02}P$, respectively **(c),(d),(e)** HARPES paths in k-space for the optimized experimental geometries determined from the FEFS calculations: GaP –red line, $Ga_{0.98}Mn_{0.02}P$ –blue line. The paths are shown in three different perspectives. $Ga_{0.98}Mn_{0.02}P$ is very close to the ideal geometry of FIG. 5. The red and blue lines overlap in **(e).**

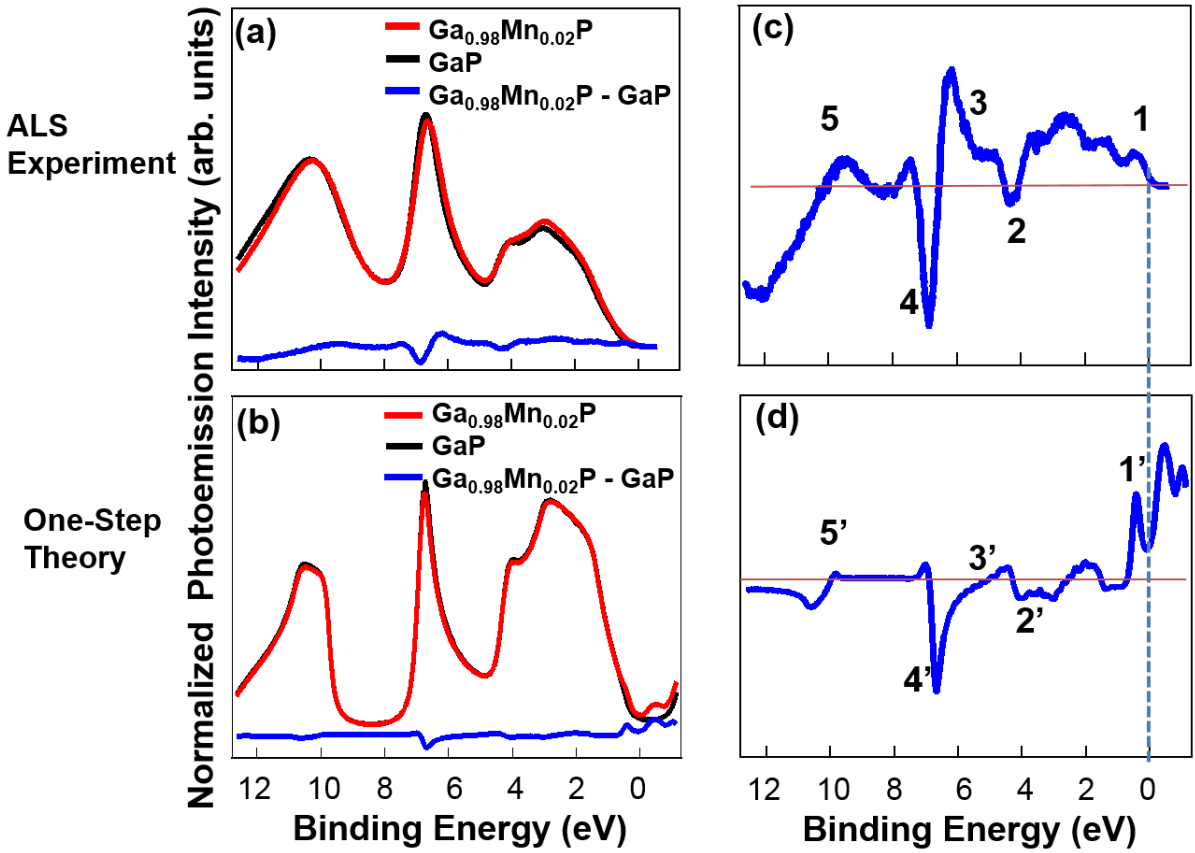

FIG. 8. **(a)** Experimental angle-integrated valence-band spectra for GaP and Ga$_{0.98}$Mn$_{0.02}$P including their difference curve, **(b)** Theoretical spectra obtained using one-step theory calculations and their difference **(c),(d),** Intensity difference spectra between Ga$_{0.98}$Mn$_{0.02}$P and GaP for the experimental data (**c**) and the one-step theory results(**d**).

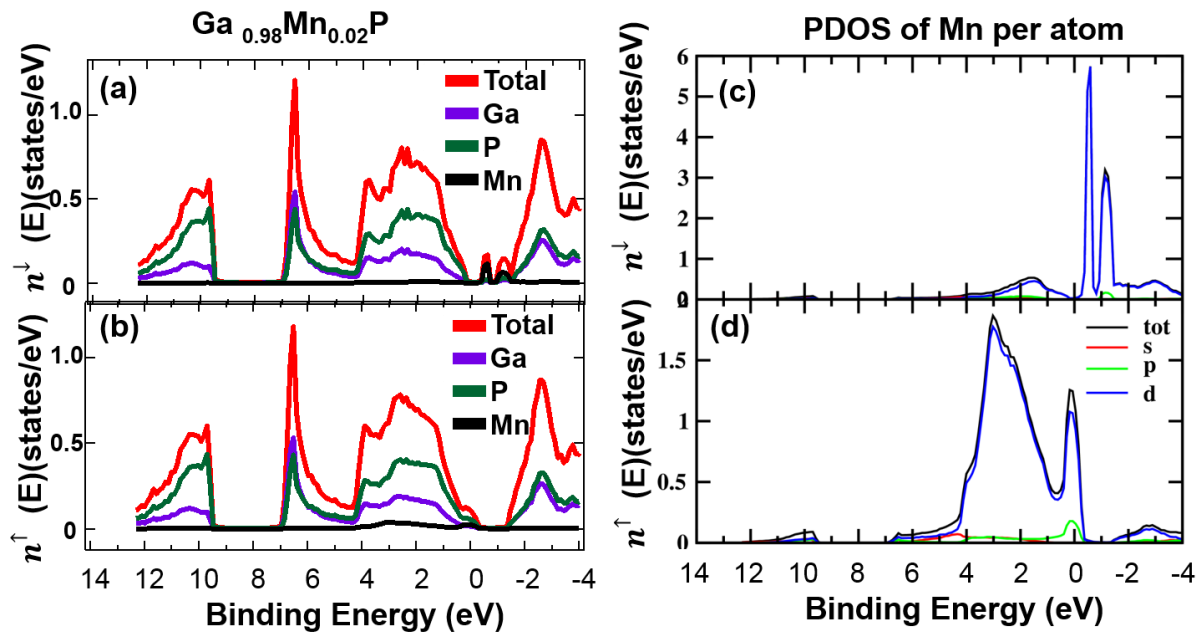

FIG. 9. Calculated spin-resolved element- and orbital-projected DOS for $Ga_{0.98}Mn_{0.02}P$, with Ga (purple), P (green) and Mn (black) contributions indicated: **(a)** = spin-down, **(b)** = spin-up. Calculated spin-resolved projected DOS for Mn alone in $Ga_{0.98}Mn_{0.02}P$, with $s$, $p$ and $d$ contributions indicated: **(c)** = spin-down, **(d)** = spin-up.

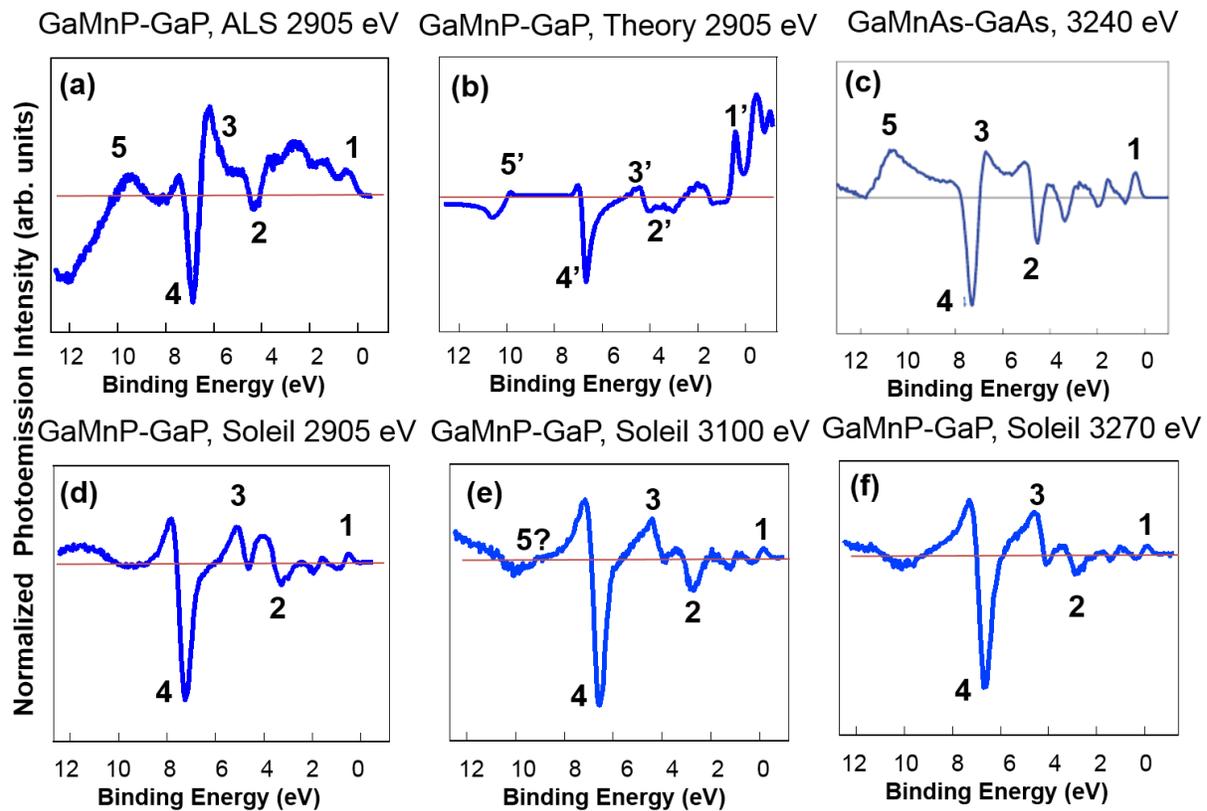

FIG. 10. Global comparison of experimental MEWDOS differences of ALS and SOLEIL data, and GaMnAs-GaAs (from ref. 9).